\documentclass[runningheads]{llncs}

\usepackage[T1]{fontenc}
\usepackage{microtype}
\usepackage{amsmath}
\usepackage{amssymb}
\usepackage{booktabs}
\usepackage{multirow}
\usepackage{graphicx}
\usepackage{xcolor}
\usepackage{url}
\usepackage{array}
\usepackage{makecell}
\title{CalBrief: A Pilot Diagnostic Benchmark for
Evidence-Calibrated Scientific Briefing with 
Large Language Models}

\titlerunning{CalBrief}

\author{Yu Fu, Yongqi Kang, Yong Zhao*}
\authorrunning{Fu et al.}
\institute{Sichuan University}

\begin{document}
\maketitle

\begin{abstract}
Large language models (LLMs) are increasingly used as research assistants, yet
it remains unclear whether they can calibrate research takeaways to the
strength and scope of the supporting evidence. We study
evidence-calibrated scientific briefing: given a bounded package of
related papers, a system should generate package-level takeaways with evidence
strength, scope boundaries, and missing-evidence caveats. We contribute a
verified pilot benchmark of 16 heterogeneous scientific evidence packages
and 96 human-verified takeaways, and we use \textsc{CalBrief}, an auditable
role/gap/strength framework, as a diagnostic probe to locate where briefing breaks down.
Under a fair-schema evaluation, structured organization improves role and gap
reasoning, but an explicit strength-calibration policy is systematically
over-conservative and falls below majority and direct-LLM baselines. To explain
why, we run a controlled diagnostic across three closed-model backbones
(GPT-4o, Claude Sonnet, Gemini Flash) that separates three potential causes of
conservatism. Approximately 63\% of the conservatism gap is attributable to
expanding the label space from binary \{moderate, weak\} to four-way
\{moderate, weak, uncertain, insufficient\_evidence\} ($p < 0.001$ across all
backbones); only $\sim$1\% is attributable to gap/scope signal injection (not
significant); the remaining $\sim$36\% arises from the pipeline policy itself.
We also find that 4-way
predictions can be post-hoc collapsed back to binary and then match or exceed
direct binary prompting, so the extra labels carry information that strict
matching hides. Label-level strength
judgment and auditable evidence organization are distinct abilities currently
in tension, and should be evaluated separately for LLM research assistants.

\keywords{Scientific document understanding \and Evidence calibration \and LLM
evaluation \and Research briefing \and Scientific summarization}
\end{abstract}

\section{Introduction}

Large language models (LLMs) are increasingly used as research assistants for
reading papers, preparing related-work summaries, and drafting technical
briefings. In realistic workflows, users often collect a small package
of related papers, such as benchmark papers, method papers, surveys, and
evaluation resources, and ask what can be concluded from the package as a
whole.

This setting raises a problem beyond ordinary summarization. A fluent summary
may still overstate what the evidence supports. Benchmark-specific gains can be
phrased as general capability claims, and an isolated method paper can be read
as field-level consensus. These are failures of evidence calibration: the
strength and scope of a generated takeaway do not match the underlying
evidence package.

We study evidence-calibrated scientific briefing: given a bounded
scientific paper package, the goal is to generate package-level takeaways
together with strength labels, evidence support, scope caveats, and
missing-evidence information. A calibrated briefing should answer not only
``what do these papers say?'', but also ``how strongly does the package
support each conclusion?'' and ``where are the limits of the evidence?''

The primary contribution is a benchmark and a diagnostic finding, not a new
system. We construct a verified pilot benchmark with 16 scientific evidence
packages and 96 human-verified package-level takeaways covering RAG
evaluation, long-context RAG, tool-use, GUI/web agents, multi-agent
evaluation, LLM safety, scientific claim verification, and scientific paper
summarization. To probe where briefing breaks down, we use
\textsc{CalBrief}, an auditable framework that decomposes briefing into
role-aware evidence organization, graph-supported gap analysis, and strength
calibration. \textsc{CalBrief} is deliberately not a new foundation model; it
is an instrument that exposes intermediate evidence signals so each
sub-ability can be measured against direct-LLM baselines.

Our findings reveal a separation between evidence organization and strength
calibration. Structured organization improves paper-role identification and
evidence-gap relevance, but its explicit strength-calibration policy is
systematically over-conservative and is beaten by the same LLMs prompted
directly. A controlled diagnostic across three closed-model backbones
attributes most of this failure to expanding the label space, finds that
adding gap and scope signals changes almost nothing, and leaves a smaller but
separable residual to the pipeline policy. A post-hoc collapse of richer
labels back to binary recovers and sometimes exceeds direct binary prompting,
showing a constructive use of label-space inflation.

\noindent\textbf{Contributions.}
\begin{itemize}
    \item We formulate evidence-calibrated scientific briefing as a
    package-level scientific document understanding task and introduce a
    human-verified pilot benchmark of 16 packages and 96 takeaways annotated
    for paper roles, evidence gaps, scope-limited claims, mismatch signals,
    and strength labels, with a double-annotation reliability check.
    \item Using \textsc{CalBrief} as a diagnostic probe, we show that
    structured organization improves role and gap reasoning, while an
    explicit calibration policy is systematically over-conservative.
    \item Across three closed-model backbones (GPT-4o, Claude Sonnet, Gemini
    Flash), we decompose the conservatism failure and attribute the bulk of it
    to label-space inflation, almost none to signal injection, and a smaller
    remainder to the pipeline policy.
    \item We show a constructive complement: post-hoc collapsing 4-way
    predictions to binary can match or exceed direct binary prompting.
\end{itemize}

\section{Related Work}

Scientific document summarization condenses one or more papers into concise
summaries~\cite{ref_scisumm,ref_multidoc}, and recent LLM systems extend this
to literature review and survey generation~\cite{ref_autosurvey}. These target
coverage and citation grounding; we instead ask whether the generated
conclusions state evidence strength and scope correctly.

Claim verification asks whether a fixed claim is supported or
refuted~\cite{ref_scifact,ref_fever,ref_claimver}, and grounded generation
requires outputs to cite their sources~\cite{ref_rag,ref_citegen}; neither
addresses the strength or scope of newly generated conclusions. We adapt the
strength, indirectness, and risk-of-bias distinctions from systematic-review
work~\cite{ref_grade,ref_systreview,ref_nlp4sr} to AI/NLP packages, where
benchmark scope and role heterogeneity dominate.

A growing body of work shows that LLMs can fail under realistic or diagnostic
settings even when they score well on standard
benchmarks~\cite{ref_diagnostic,ref_overclaim,ref_overclaim_1,ref_calibration,ref_calibration_1}.
We follow this tradition but attribute the failure to separable sources
(label space, signals, policy) rather than a single capability gap.

\section{Task and Verified Pilot Benchmark}
\label{sec:task}

\subsection{Task Definition}

The input is a bounded scientific paper package $P = \{p_1, p_2, \ldots,
p_n\}$, where each paper contributes a different form of evidence. A package
is associated with a research question. Open-ended paper retrieval is
outside the scope of this work. The desired output is a set of package-level
calibrated takeaways $T = \{(t_i, y_i, s_i)\}_{i=1}^{m}$, where $t_i$ is a
takeaway, $y_i$ is a strength label, and $s_i$ contains evidence support,
scope caveats, and supporting paper or claim identifiers. In our verified
set, gold strength labels are moderate and weak. A moderate
takeaway is directly supported by benchmark evidence or multiple consistent
sources; a weak takeaway is only partially, indirectly, or narrowly
supported.

\subsection{Evidence Units and Dataset Construction}
\label{sec:evidence_units}

We annotate five types of evidence units: paper roles, evidence gaps,
scope-limited claims, benchmark/capability mismatch signals, and calibrated
takeaways. Paper roles include conceptual synthesis, empirical method,
evaluation resource, aggregate evidence, application/deployment,
position/perspective, and other. The annotation guideline defines a key
calibration rule that anchors all gold strength labels: scope
limitation alone does not downgrade a takeaway. A benchmark-scoped
conclusion is gold moderate when (i) the wording stays within the
measured scope and (ii) the in-scope evidence directly supports it; it is weak only when support is partial, indirect, or contested even within
that scope. A takeaway is never downgraded merely because the package lacks
deployment evidence, external validation, or cross-benchmark aggregation.
This rule is precisely the distinction the calibration probe later fails to
preserve (Section~\ref{sec:error}).

The benchmark is constructed through a semi-automated but human-controlled
workflow. Candidate papers are retrieved and shortlisted per package; only
shortlisted open-access PDFs are parsed. Draft metadata, evidence audits,
and takeaways are generated, and an LLM-assisted cleaning layer flags
wrong-package wording, missing PDFs, unsupported takeaways, role
inconsistencies, and suspicious aggregate-evidence assignments. No LLM draft
is auto-promoted to gold; each package passes a human verification checklist
and JSON write-back stage. The 16 packages span five domain groups: RAG /
retrieval evaluation (4 packages), LLM-agent benchmarks (6), scientific NLP
tasks (3), safety / trustworthiness (2), and other multi-document QA (1),
covering empirical method, evaluation resource, and conceptual synthesis as
the dominant evidence roles.

We deliberately kept this first release small and used binary gold labels. Our
goal here was a clean, fully human-verified signal rather than scale: with 96
takeaways we could check every label by hand, and the low agreement we later
report on fine-grained mismatch typing (Section~\ref{sec:task}) convinced us
that pushing for a larger or finer-grained gold set would have mixed two
different problems. We would rather report a small set we trust than a large
set we cannot stand behind.

\begin{table}[!t]
\centering
\small
\setlength{\tabcolsep}{6pt}
\begin{tabular}{lrl}
\toprule
Dataset component & Count & Notes \\
\midrule
Verified packages & 16 & Heterogeneous scientific evidence packages \\
Verified takeaways & 96 & 61 moderate / 35 weak \\
IAA subset items & 30 & Stratified across six packages \\
IAA packages & 6 & Role, strength, gap, boundary/mismatch units \\
\bottomrule
\end{tabular}
\caption{Summary of the verified pilot benchmark and reliability subset.}
\label{tab:dataset_summary}
\end{table}

\subsection{Human Double-Annotation Check}

We run a human double-annotation check on a 30-item stratified subset covering
six packages (Table~\ref{tab:iaa}). Overall exact agreement is 80.0\%.
Agreement is highest for takeaway strength (accuracy $=1.000$, Cohen's
$\kappa=1.000$) and evidence-gap relevance (accuracy $=0.833$,
$\kappa=0.667$). Paper-role labels show moderate agreement ($0.667$ /
$0.400$); fine-grained benchmark-mismatch typing has low agreement ($0.250$
/ $0.200$) and is used only diagnostically. The strength-agreement result
should be interpreted cautiously because it covers only 12 double-annotated
strength items; even so, it provides preliminary evidence that the
binary strength labels are consistently interpretable.

\begin{table}[!htbp]
\centering
\small
\setlength{\tabcolsep}{5pt}
\begin{tabular}{lrrr}
\toprule
Annotation type & $n$ & Accuracy & $\kappa$ \\
\midrule
Paper role label & 6 & 0.667 & 0.400 \\
Takeaway strength & 12 & 1.000 & 1.000 \\
Evidence gap relevance & 6 & 0.833 & 0.667 \\
Scope boundary judgment & 2 & 1.000 & 1.000 \\
Benchmark mismatch type & 4 & 0.250 & 0.200 \\
\midrule
Overall & 30 & 0.800 & -- \\
\bottomrule
\end{tabular}
\caption{Human double-annotation agreement on a 30-item stratified subset.}
\label{tab:iaa}
\end{table}

\section{CalBrief Framework}
\label{sec:framework}

\textsc{CalBrief} is an auditable probe with three components whose purpose
is not fluency but exposing the evidential basis and scope of each
conclusion, so that each sub-ability can be measured separately.
\textbf{Role-aware evidence organization} identifies paper roles
before producing takeaways, to reduce evidence-role confusion.
\textbf{Graph-supported gap analysis} organizes papers, roles, claims,
gaps, benchmark coverage, and scope-limited claims into an inspectable
structure; the graph is an auditable representation for evidence analysis,
not a learned model. \textbf{Strength calibration} maps support, benchmark
coverage, scope boundaries, deployment gaps, and role heterogeneity to a
strength label and revises wording to match. As the experiments show, this
last step is the hardest part and is prone to over-conservatism.

The calibration step itself is a deterministic policy applied after LLM
extraction of evidence units. In compact form, given supporting papers
$\mathit{Sup}(t)$, roles $R(t)$, typed gaps $G(t)$, and mismatch signals
$M(t)$ for takeaway $t$, return \textsc{insufficient\_evidence} if
$|\mathit{Sup}(t)|=0$; \textsc{uncertain} if any $g \in G(t)$ matches
$t$'s scope (external\_validation or capability\_coverage) or any
$m \in M(t)$ is unresolved; \textsc{moderate} if $|R(t)| \geq 2$ are
complementary and no high-severity $g$ remains; otherwise \textsc{weak}.
Edge cases follow the same priority order, and the executable policy will be
released with the benchmark artifacts. This policy is what we later
attribute the residual conservatism gap to in
Section~\ref{sec:diagnostic}.

\section{Experimental Setup}

\paragraph{Methods and fair-schema evaluation.}
For the main structured comparison we evaluate a strong structured prompt,
role-agnostic, role-aware, and role-aware graph pipelines, and
\textsc{CalBrief}; an always-moderate majority baseline and direct briefing
baselines using Kimi-K2.6, GLM-5.1, Qwen3-32B serve as strength-only
references. All structured methods output the same fields (paper roles,
evidence gaps, calibrated conclusions with strength labels,
unsupported-inference flags, scope caveats) and produce parseable outputs
for all 16 packages. Direct-briefing baselines do not emit role/gap
structure; their role and gap entries are marked \emph{N/A} rather than
treated as incorrect.

\paragraph{Closed-model diagnostic.}
To separate the causes of \textsc{CalBrief}'s strength-calibration failure,
we run a controlled diagnostic over the same 96 takeaways using GPT-4o,
Claude Sonnet, and Gemini Flash. We picked these three because they are
widely used closed models from three different providers; if the same
conservatism shows up in all of them, we can be more confident it is a
property of the task setup rather than a quirk of one model family. We compare three conditions per backbone: direct binary (label space \{moderate, weak\}, no pipeline signals); direct 4-way clean (expanded label space \{moderate, weak, uncertain,
insufficient\_evidence\}, no pipeline signals); and direct 4-way with
signals (4-way labels plus the gap, scope, and benchmark-coverage signals
\textsc{CalBrief} would inject). Each condition is run under two prompt
versions: \texttt{v1\_unbiased} (neutral definitions) and
\texttt{v2\_calbrief\_aligned} (\textsc{CalBrief}'s label definitions
verbatim). Significance is assessed by paired bootstrap with 10{,}000
resamples. All backbones produce 96/96 valid outputs after
\texttt{--resume --retry-errors}.

\paragraph{Metrics.}
We report role accuracy, evidence-gap relevance, strength exact match,
strength macro-F1 over \{moderate, weak\}, and unsupported inference. For 4-way conditions we additionally report conservative collapse
(\{uncertain, insufficient\_evidence\} $\to$ weak) and
aggressive collapse (\{uncertain\} $\to$ moderate;
\{insufficient\_evidence\} $\to$ weak). Benchmark-mismatch detection
is excluded from the main table due to low human agreement.

\section{Results}
\label{sec:results}

\subsection{Main Fair-Schema Results}

\begin{table}[!htbp]
\centering
\small
\setlength{\tabcolsep}{4pt}
\renewcommand{\arraystretch}{1.05}
\resizebox{\linewidth}{!}{%
\begin{tabular}{@{}lccccc@{}}
\toprule
Method & Role Acc. & Gap Rel. & Str. Acc. & Str. F1 & Unsup. \\
\midrule
Strong structured prompt & 0.0000 & 0.3750 & 0.0000 & 0.0000 & 0.0000 \\
Role-agnostic pipeline & 0.0000 & 0.3681 & 0.0000 & 0.0000 & 0.0000 \\
Role-aware pipeline & 0.4607 & 0.3574 & 0.0000 & 0.0000 & 0.0000 \\
Role-aware graph & \textbf{0.4607} & \textbf{0.4115} & 0.0000 & 0.0000 & 0.0000 \\
\textsc{CalBrief} & \textbf{0.4607} & \textbf{0.4115} & 0.0313 & 0.0406 & 0.0000 \\
\midrule
Always-moderate & N/A & N/A & 0.6354 & 0.3867 & N/A \\
Direct briefing, Kimi-K2.6 & N/A & N/A & \textbf{0.6875} & 0.6191 & N/A \\
Direct briefing, GLM-5.1 & N/A & N/A & 0.6458 & \textbf{0.6250} & N/A \\
Direct briefing, Qwen3-32B & N/A & N/A & 0.6354 & 0.4431 & N/A \\
\bottomrule
\end{tabular}%
}
\caption{Main verified evaluation. Structured rows are evaluated under the
fair schema. Direct-briefing and majority rows are strength-only baselines.}
\label{tab:fair_schema_main}
\end{table}

Table~\ref{tab:fair_schema_main} shows that evidence organization and
strength calibration behave differently. Among structured methods,
role-aware modeling raises role accuracy from 0 to 0.4607 and graph support
raises gap relevance from 0.3574 to 0.4115. We read 0.4607 not as a
reliable role labeler, but as evidence that explicit role modeling is useful
while still insufficient: the main residual errors are confusions between
adjacent roles (survey vs.\ aggregate evidence; empirical method vs.\
evaluation resource), which are exactly the distinctions a briefing system
must not collapse. Structured pipelines therefore offer inspectable
organization that is not yet accurate, in contrast to direct briefing, which
offers no organization at all.

Strength calibration behaves in the opposite way. \textsc{CalBrief} obtains
only 0.0313 / 0.0406 (exact match / macro-F1), far below the always-moderate
baseline (0.6354 / 0.3867). A diagnostic ScopeAware variant also fails
(0.0104 / 0.0125), which indicates that a simple rule forbidding scope-driven
downgrades is not enough on its own.

\subsection{Direct Briefing: Strong Labels, No Auditability}

The same three backbones, prompted directly, preserve a much stronger
moderate/weak signal: Kimi-K2.6 0.6875/0.6191, GLM-5.1 0.6458/0.6250,
Qwen3-32B 0.6354/0.4431 (exact match / macro-F1), all above always-moderate
in macro-F1. But the advantage does not extend to auditable organization:
direct briefing emits no role/gap structure. These two observations describe
a trade-off that is itself part of our finding: direct LLMs give better
labels, while structured pipelines give inspectable evidence. When the same
LLMs are forced to emit the full fair schema, their strength collapses to the
always-moderate prior, so full-schema direct prompting is not a solution
either.

Post-hoc binary collapse of \textsc{CalBrief}'s 4-way predictions raises macro-F1 to 0.3031 (conservative) or 0.5286 (aggressive); the near-universal downgrading behind these labels is examined in Section~\ref{sec:error}.

\subsection{Diagnostic Decomposition: Label Space, Signals, Policy}
\label{sec:diagnostic}

The results above show that \textsc{CalBrief}'s strength calibration
fails, but not why. The pipeline differs from a direct binary prompt
in three ways that could each contribute to over-conservatism: (i) it
predicts in a four-way label space; (ii) it injects gap, scope, and
benchmark-coverage signals; and (iii) it executes an explicit calibration
policy. To isolate these factors, we ran the diagnostic of Section~5 over
the same 96 takeaways using GPT-4o, Claude Sonnet, and Gemini Flash. We
think the decomposition matters more than the single gap number: a designer
who only sees that CalBrief underperforms might try to fix the policy, when
in fact most of the damage is done before the policy ever runs.

\paragraph{Per-backbone results.}
Table~\ref{tab:closed_model_4way} reports strict Macro-F1 for every
backbone, condition, and prompt version, and the pattern is the same across all of them: direct binary prompting is strongest (in line with the open-model baselines of Table~\ref{tab:fair_schema_main}), moving to a 4-way label space sharply lowers it, adding pipeline signals barely changes it, and the full \textsc{CalBrief} pipeline (0.0406) falls below every direct 4-way cell.

\begin{table}[!t]
\centering
\small
\setlength{\tabcolsep}{4pt}
\renewcommand{\arraystretch}{1.05}
\resizebox{\linewidth}{!}{%
\begin{tabular}{@{}llccc@{}}
\toprule
Condition & Prompt & Claude Sonnet & GPT-4o & Gemini Flash \\
\midrule
Direct binary & -- & 0.5553 & \textbf{0.6221} & 0.5501 \\
\midrule
Direct 4-way clean & v1 & 0.2389 & \textbf{0.2874} & 0.2232 \\
Direct 4-way clean & v2 & \textbf{0.2561} & 0.2534 & 0.2168 \\
Direct 4-way w/ signals & v1 & 0.2135 & \textbf{0.2616} & 0.2200 \\
Direct 4-way w/ signals & v2 & \textbf{0.2700} & 0.2708 & 0.1968 \\
\midrule
\textsc{CalBrief} strict & -- & \multicolumn{3}{c}{0.0406 (pipeline-level)} \\
\bottomrule
\end{tabular}%
}
\caption{Strict Macro-F1 across three backbones, three conditions, and two
prompt versions. All cells have 96/96 valid predictions. Macro-F1 here is pooled (one F1 over all 96 takeaways per cell), whereas
Table~\ref{tab:three_component} and Figure~\ref{fig:three_component} report per-package mean Macro-F1 (F1 per package, averaged across the 16
packages); bootstrap resampling for the decomposition is performed at the
package level ($n{=}16$), with cells treated as paired observations.}
\label{tab:closed_model_4way}
\end{table}

\begin{figure}[!t]
\centering
\includegraphics[width=0.9\linewidth]{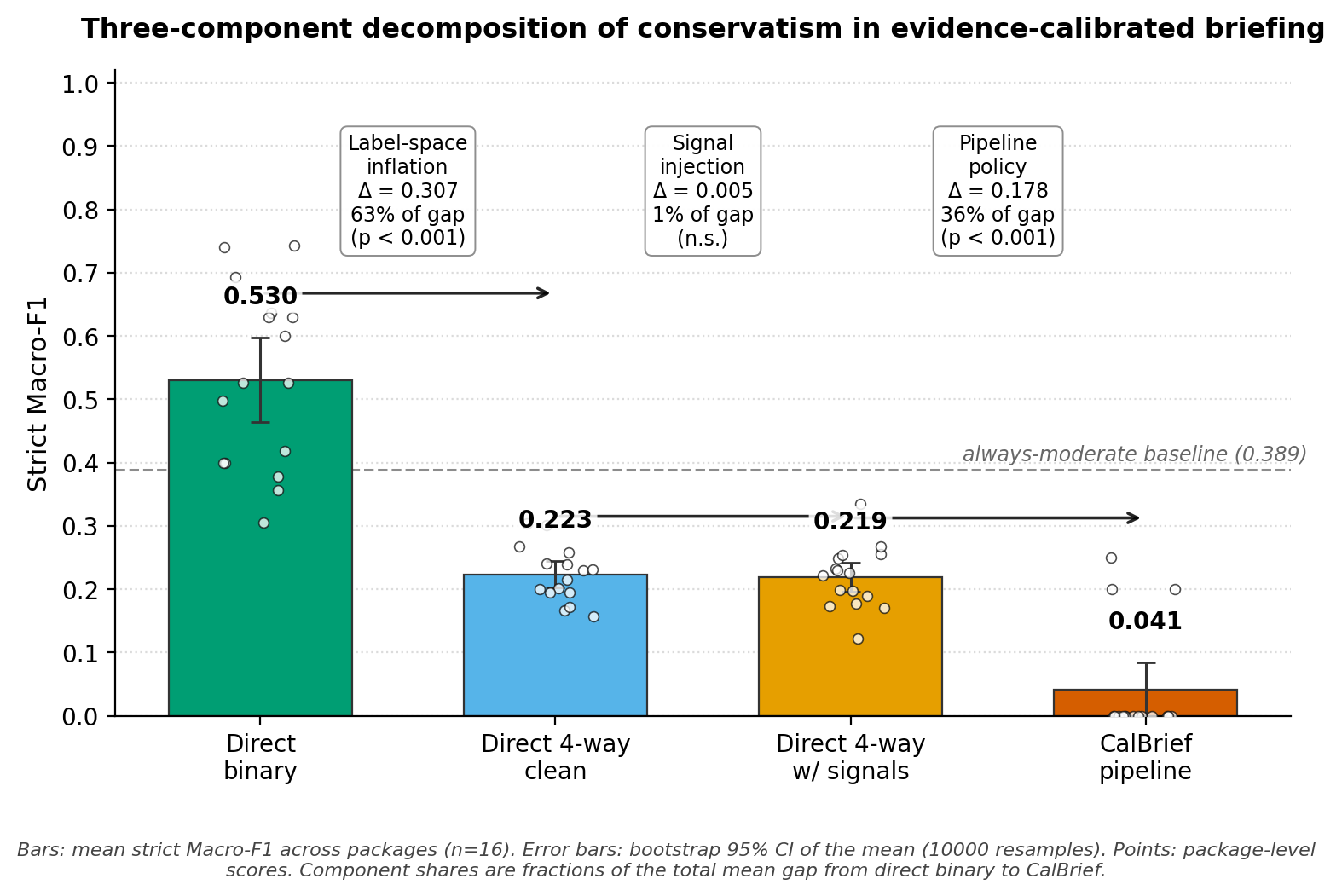}
\caption{Three-component decomposition of conservatism. Bars: mean
per-package strict Macro-F1 (averaged across packages, then across cells);
error bars: bootstrap 95\% CI of the mean (10{,}000 resamples); points:
individual cells. Binary $\to$ 4-way clean accounts for $\sim$63\% of the
0.489 gap ($p < 0.001$); clean $\to$ with-signals contributes $\sim$1\%
(n.s.); the remaining $\sim$36\% is attributable to the \textsc{CalBrief}
pipeline.}
\label{fig:three_component}
\end{figure}

\begin{table}[!t]
\centering
\small
\setlength{\tabcolsep}{5pt}
\renewcommand{\arraystretch}{0.95}
\begin{tabular}{lccc}
\toprule
Condition & Mean strict & $\Delta$ vs.\ & Share of \\
 & Macro-F1 & previous row & total gap \\
\midrule
Direct binary & 0.530 & -- & -- \\
Direct 4-way clean & 0.223 & $-0.307$ ($p < 0.001$) & \textbf{63\%} \\
Direct 4-way w/ signals & 0.219 & $-0.005$ (n.s., $p>0.18$) & \textbf{1\%} \\
\textsc{CalBrief} pipeline & 0.041 & $-0.178$ & \textbf{36\%} \\
\bottomrule
\end{tabular}
\caption{Three-component decomposition. Means are over per-package strict
Macro-F1, averaged across packages and across the three backbones and two
prompt versions. $\Delta$ values are absolute drops from the previous row;
significance from paired bootstrap with 10{,}000 resamples. Total gap from
direct binary to \textsc{CalBrief}: 0.489 Macro-F1. Component shares are
rounded.}
\label{tab:three_component}
\end{table}

\paragraph{Three-component decomposition.}
Figure~\ref{fig:three_component} and Table~\ref{tab:three_component}
report averages over backbones and prompts. Moving from binary to clean
4-way drops strict Macro-F1 by 0.307 on average (paired bootstrap $p <
0.001$ for every backbone and prompt version), accounting for $\sim$63\% of
the total 0.489 gap. Adding gap and scope signals produces a further drop of
only 0.005 on average, not significant ($p > 0.18$). The remaining 0.178
absolute gap (36\%) is attributable to pipeline-internal calibration logic
that combines roles, gaps, and scope into a final label. The bulk of
\textsc{CalBrief}'s strict-label degradation is therefore not explained by
signal injection, and is largely induced by the expanded label
space under strict binary matching. Three strong, modern LLMs are
substantially more conservative under strict binary matching once they are
given the option to say uncertain or insufficient\_evidence, even when no policy and no signals are
present. The signals \textsc{CalBrief} computes are not, by
themselves, the proximate cause of over-conservatism.

\paragraph{Post-hoc collapse to binary.}
\label{sec:collapse}
Because our gold labels are binary, the $\sim$63\% label-space component measures degradation under strict binary matching rather than a proven loss of epistemic calibration: a model that hedges to uncertain on a genuinely thin case is scored the same as one that hedges arbitrarily. The post-hoc collapse makes this concrete. The same 4-way predictions can be mapped to binary (\{uncertain, insufficient\_evidence\} $\to$ weak). For Claude
Sonnet this raises Macro-F1 from 0.270 (strict) to 0.647 on the
best cell (4-way with signals, v2), exceeding direct binary
prompting (0.555) by $+0.092$ on the same items; every Claude 4-way
condition exceeds direct binary after collapse. GPT-4o and Gemini show no
such gain (best collapsed F1: 0.608 vs.\ binary 0.622; 0.513 vs.\ 0.550
respectively). The richer label set therefore does not just add noise; it can
hold on to information that a principled mapping recovers. Reading this
together with the per-backbone numbers, we suspect future systems will do
better by predicting in a richer space and collapsing afterward than by
forcing a flat binary judgment up front, though the effect we see is clearest
on Claude and we would not over-generalize from one model. We report the
strongest and most representative collapsed results here; full prediction logs
will be released with the benchmark artifacts.

\section{Error Analysis and Discussion}
\label{sec:error}

\subsection{Systematic Conservatism in Explicit Calibration}

The dominant strength error is directional conservatism. Of the 61
gold-moderate takeaways, only 2 are predicted moderate; the remaining 59
are downgraded (22 to weak, 16 to uncertain, 21 to insufficient evidence).
No gold-weak takeaway is upgraded to moderate, and within the weak class
33 of 35 are pushed to insufficient evidence. Every class is shifted away
from moderate and toward evidence-insufficiency, so the shift is systematic
rather than random. All methods see identical parsed inputs, and all 96
takeaways were assigned moderate or weak labels through human verification,
so the insufficiency labels reflect the disposition of the calibration policy
rather than missing gold annotations for the evaluated items.

A recurring case-level pattern is that the policy treats missing higher-level
evidence as a reason to weaken even properly bounded takeaways. In
benchmark-centered packages a takeaway may be directly supported within the
dataset scope, yet the policy penalizes it because the package lacks
deployment evidence, external validation, or cross-benchmark aggregation.
This confuses two notions: a claim can be bounded without being
unsupported. A benchmark-scoped conclusion should not be upgraded to
a deployment claim, but neither should it be downgraded to insufficient
evidence when the wording already respects the measured scope, which is
precisely the gold rule of Section~\ref{sec:evidence_units}. Direct
briefing shows the opposite tendency: it more readily assigns moderate when
the takeaway is semantically supported, explaining its stronger macro-F1,
but it does not organize roles, gaps, or scope, so its labels are less
auditable.

\subsection{Interaction Between Label Wording and Signal Injection}

The clean vs.\ with-signals comparison exposes an interaction between
label-definition wording and signal injection. Under \texttt{v1\_unbiased},
where definitions do not explicitly reference benchmark scope or thin
evidence, adding signals slightly hurts strict F1 across all three
models (Claude $-0.025$, GPT-4o $-0.026$, Gemini $-0.003$). Under
\texttt{v2\_calbrief\_aligned}, where definitions explicitly mention
``benchmark settings'' and ``mismatched evidence'', the same signals
slightly help on average (Claude $+0.014$, GPT-4o $+0.017$; Gemini
still $-0.020$). Individual effects are small and not significant per cell,
but the directional pattern indicates that \textsc{CalBrief}'s signals and
label semantics are co-adapted: signals about gaps and scope only become
useful in the presence of definitions that explicitly reference them.
Designing signals and labels independently, as is common when assembling
modular calibration pipelines, may therefore fail to realize the intended
calibration effect.

\subsection{Implications and Limitations}

Two practical implications follow, and we state them as the lessons we drew
rather than as established rules. First, label spaces should be designed
jointly with the calibration policy and signals; in our runs, simply adding
uncertain or insufficient\_evidence as options, without label definitions
tuned to them, was enough to make the models hedge. Second, when
richer labels are needed for auditability, the binary view should be
recovered through post-hoc collapse rather than by forcing prediction in a
flat space up front.

\textbf{Limitations.} This is a verified pilot (16 packages, 96 takeaways,
binary gold). Role accuracy is far from saturated (0.4607), so we report
structured organization as inspectable rather than reliable, and fine-grained
mismatch typing is used only diagnostically due to low IAA. The collapse
finding is more model-dependent (most pronounced on Claude). A
natural extension is to re-annotate a subset with native 4-way labels, which
we did not pursue here because IAA on fine-grained mismatch is already low
($\kappa{=}0.20$), so reliable 4-way annotation is a separate
undertaking. Because briefing systems influence literature reviews and
planning, they should expose evidence boundaries rather than replace expert
judgment in high-stakes domains. Annotation guidelines, package lists,
prompts, model versions, decoding settings, full prediction logs, and the
executable \textsc{CalBrief} policy will be released with the benchmark.

\section{Conclusion}

We introduced evidence-calibrated scientific briefing and a pilot benchmark
of 16 packages and 96 takeaways, using \textsc{CalBrief} as an auditable
probe. Structured organization yields inspectable paper-role and gap fields
that direct briefing lacks, but its explicit strength calibration is
systematically over-conservative and is beaten by the same LLMs prompted
directly. A closed-model diagnostic decomposes the failure: the expanded
label space accounts for the bulk of the degradation, signal injection
alone contributes essentially nothing, and the pipeline policy adds a
smaller but separable additional cost. Post-hoc collapsing 4-way predictions
to binary can match or exceed direct binary prompting. Label-level strength
judgment and auditable evidence organization should be evaluated as distinct
dimensions for LLM research assistants, and we will release the benchmark to
enable this.

\bibliographystyle{splncs04}
\bibliography{reference}

\end{document}